\documentclass[%
reprint,
superscriptaddress,
 amsmath,amssymb,
 aps,
]{revtex4-2}

\usepackage{color}
\usepackage{soul}
\usepackage{CJK}
\usepackage{graphicx}%
\usepackage{dcolumn}%
\usepackage{bm}%
\usepackage{hyperref}%

\def\dd{\mathrm{d}}
\def\ee{\mathrm{e}}

\CJKfamily{gbsn}

\begin{document}

\begin{CJK*}{UTF8}{gbsn}

\title{Limits on the accuracy of contact inhibition of locomotion}%

\author{Wei Wang (汪巍)}
\affiliation{%
Department of Physics and Astronomy, Johns Hopkins University, Baltimore, Maryland 21218, USA\\
}%

\author{Brian A. Camley}%
\affiliation{%
Department of Physics and Astronomy, Johns Hopkins University, Baltimore, Maryland 21218, USA\\
}%
\affiliation{%
Department of Biophysics, Johns Hopkins University, Baltimore, Maryland 21218, USA\\
}

\begin{abstract}
Cells that collide with each other repolarize away from contact, in a process called contact inhibition of locomotion (CIL), which is necessary for correct development of the embryo. CIL can occur even when cells make a micron-scale contact with a neighbor -- much smaller than their size. How precisely can a cell sense cell-cell contact and repolarize in the correct direction? What factors control whether a cell recognizes it has contacted a neighbor?
We propose a theoretical model for the limits of CIL where cells recognize the presence of another cell by binding the protein ephrin with the Eph receptor. This recognition is made difficult by the presence of interfering ligands that bind nonspecifically. %
Both theoretical predictions and simulation results show that it becomes more difficult to sense cell-cell contact when it is difficult to distinguish ephrin from the interfering ligands, or when there are more interfering ligands, or when the contact width decreases. However, the error of estimating contact position remains almost constant when the contact width changes. This happens because the cell gains spatial information largely from the boundaries of cell-cell contact.
We study using statistical decision theory the likelihood of a false positive CIL event in the absence of cell-cell contact, and the likelihood of a false negative where CIL does not occur when another cell is present. Our results suggest that the cell is more likely to make incorrect decisions when the contact width is very small or so large that it nears the cell's perimeter. However, in general, we find that cells have the ability to make reasonably reliable CIL decisions even for very narrow (micron-scale) contacts, even if the concentration of interfering ligands is ten times that of the correct ligands. %
\end{abstract}

\maketitle
\end{CJK*}

\section{\label{sec:intro}Introduction}

More than half a century ago, researchers first observed contact inhibition of locomotion (CIL), when two colliding fibroblasts changed direction rapidly and migrated away from the collision~\cite{CIL, PhysRevE.106.054413, singh2021rules}. CIL is a well-known characteristic of normal cells and plays an important role in regulating cell motility~\cite{singh2021rules, PhysRevE.106.054413}, tissue growth~\cite{zimmermann2016contact}, and development~\cite{larranaga_ephrin_nodate, lin_interplay_2015, exp}. For example, experiments revealed that \emph{Drosophila} macrophages (haemocytes) require CIL for their uniform embryonic dispersal~\cite{exp, stramer2010clasp}. {CIL can happen -- though rarely -- even when cells are crawling on isolated nanofibers, leading to very small ($\sim \mathrm{\mu m}$) contact regions \cite{singh2021rules}.}
How can cells reliably observe the presence of another cell just by such a small contact?

CIL is regulated by the Eph-ephrin signaling pathway~\cite{exp, larranaga_ephrin_nodate, zapata-mercado_efficacy_2022, lisabeth2013eph}. Contact detected by Eph-ephrin or cadherin binding \cite{roycroft-mayor-review, noordstra2021cadherin} reorganizes the actin cytoskeleton~\cite{klein2012eph} by modulating the main elements of the Rho GTPase family~\cite{noren2004eph, arvanitis2008eph}, resulting in the increase of actomyosin contractility and the inhibition of lamellipodia formation at the cell-cell contact~\cite{pasquale2008eph, larranaga_ephrin_nodate}.
Eph receptors are a family of transmembrane receptor tyrosine kinases (RTK)~\cite{zapata-mercado_efficacy_2022, larranaga_ephrin_nodate} that  regulate tissue boundary formation in development~\cite{cayuso2015mechanisms} and tissue organization maintenance in adult organisms~\cite{larranaga_ephrin_nodate}. Their ligands, known as ephrins, are normally anchored to the cell surface~\cite{zapata-mercado_efficacy_2022, larranaga_ephrin_nodate, lin_interplay_2015} -- Eph-ephrin signaling occurs when cells come into contact with one another. Ephrin cues can be used to reorganize and repolarize cells. Ephrin-coated beads lead to CIL when cells collide with them \cite{lin_interplay_2015}, and micropatterned ephrin spatial cues can be used to align intestinal crypt cells~\cite{larranaga_ephrin_nodate}.

Cells can use receptors on their membrane to estimate external ligand concentrations~\cite{PhysRevLett.124.028101, PhysRevE.99.022423, chakrabortty2019theoretical,BergPurcell} and also to sense spatial~\cite{PhysRevE.83.021917, PhysRevE.100.022401, PhysRevResearch.2.043146, PhysRevE.105.044410, PhysRevLett.105.048104,nwogbaga2022physical} or temporal~\cite{PhysRevLett.104.248101} gradients. The accuracy limits of these estimations have been extensively studied~\cite{ PhysRevE.99.022423, chakrabortty2019theoretical, PhysRevE.83.021917, PhysRevE.100.022401, PhysRevResearch.2.043146, PhysRevE.105.044410}, including related measurements of spatially localized extracellular signals~\cite{PhysRevE.100.022401}.
{For a cell to undergo CIL, it must detect the presence and location of the cell-cell contact -- akin to measuring a spatially localized chemical.} {The cell's ability to make this detection will be challenged if ligands other than ephrin, i.e., spurious ligands, bind non-specifically to the Eph receptors~\cite{mora, singh_simple_2017, feinerman2008quantitative, chylek2014interaction}. Potential candidates for the spurious ligands are soluble ephrins \cite{wykosky2008soluble},  ephrins on extracellular vesicles \cite{gong2016exosomes}, or ligands for other RTKs, e.g., growth factors~\cite{minami2011ephrina}.} In scenarios where there is only a small patch of contact \cite{singh2021rules}, such interference from other ligands poses a significant threat to the precision of cell sensing.

In this paper, we develop a theory for sensing cell-cell contact {via Eph receptor-ephrin contact in the presence of spurious ligands} using a multi-ligand model~\cite{mora, singh_simple_2017, PhysRevE.99.022423, lalanne2015chemodetection}.
To analyze the optimal performance of such a sensing mechanism, we adopt the method of maximum likelihood estimation (MLE) and derive the fundamental sensing limits by using the Cram\'er-Rao bound. {We find that the accuracy of detecting the location of the other cell is degraded when the binding affinity of the spurious ligands is close to the true ephrin's affinity, or when the fraction of correct ligands is small. Surprisingly, accuracy does not depend strongly on the contact width. These results are also supported by Monte Carlo simulations.} %
We use statistical decision theory to study the problem of whether the cell can effectively detect the presence of another cell via a small cell-cell contact, {and show how cells can trade off false negatives, where they fail to respond to another cell, and false positives, where they react even in the absence of a contacting cell. We find that cell-cell contact can be highly reliably detected even at large levels of spurious ligand concentrations.}

\section{Model}
\begin{figure}
\includegraphics[width=0.48\textwidth]{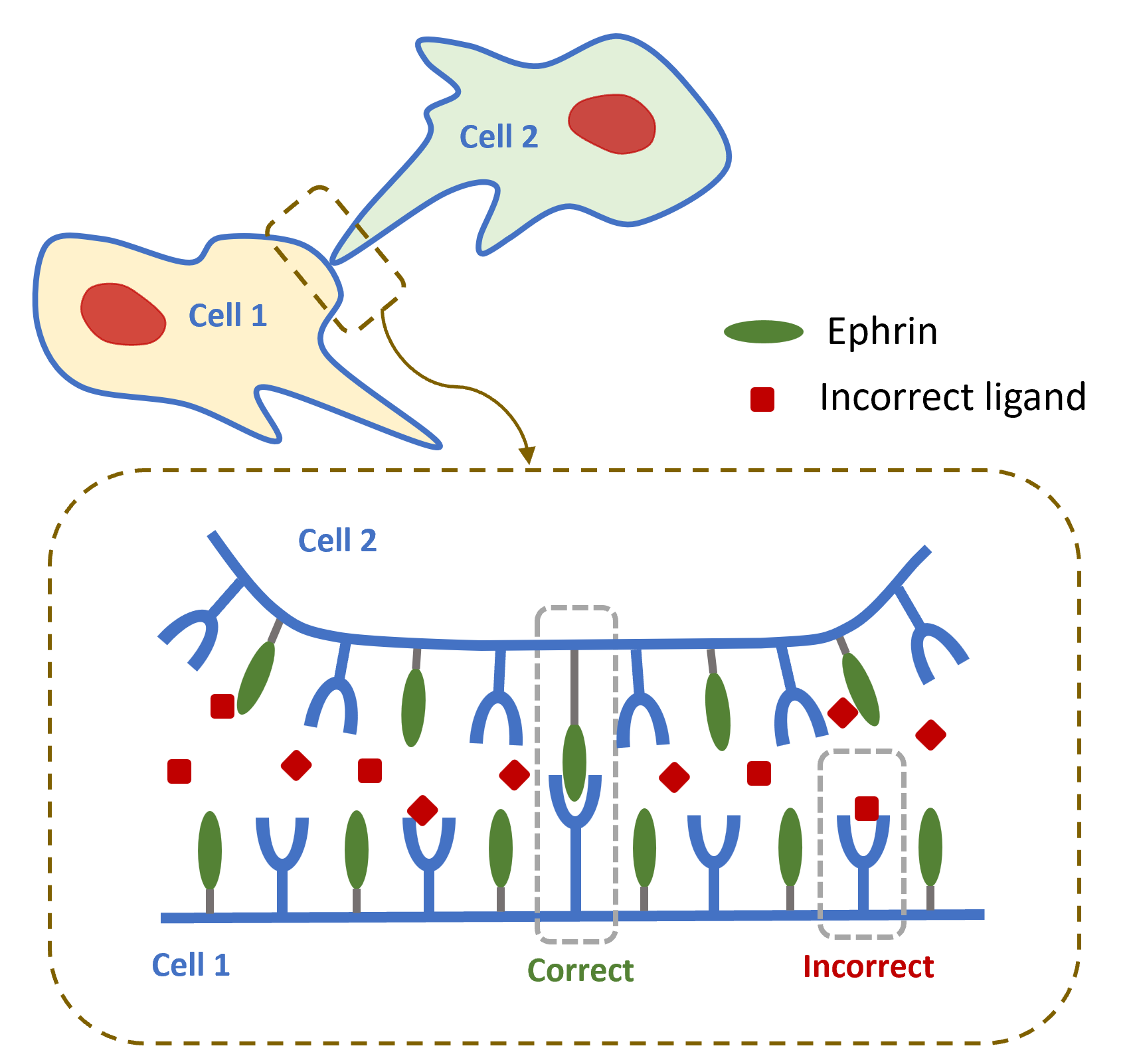}%
\caption{\label{fig:schematics} Illustration of two cells in contact where they can sense each other through Eph-ephrin signaling. Both Eph receptors (blue receptors) and their ligands - ephrins (green ellipses), are distributed on the cell membranes. Besides the cognate ligands, other spurious ligands (red squares) can also bind to the Eph receptors. These incorrect binding events mixed with the correct Eph-ephrin binding will affect the sensing accuracy of contact inhibition of locomotion. }
\end{figure}

Consider two cells coming into contact as shown in Fig. \ref{fig:schematics}, where both cells can sense each other through Eph-ephrin binding \cite{larranaga_ephrin_nodate,lin_interplay_2015}. We {study the sensing accuracy for cell $1$ detecting cell $2$, as the mirror problem is equivalent.}

\begin{figure}
\includegraphics[width=0.48\textwidth]{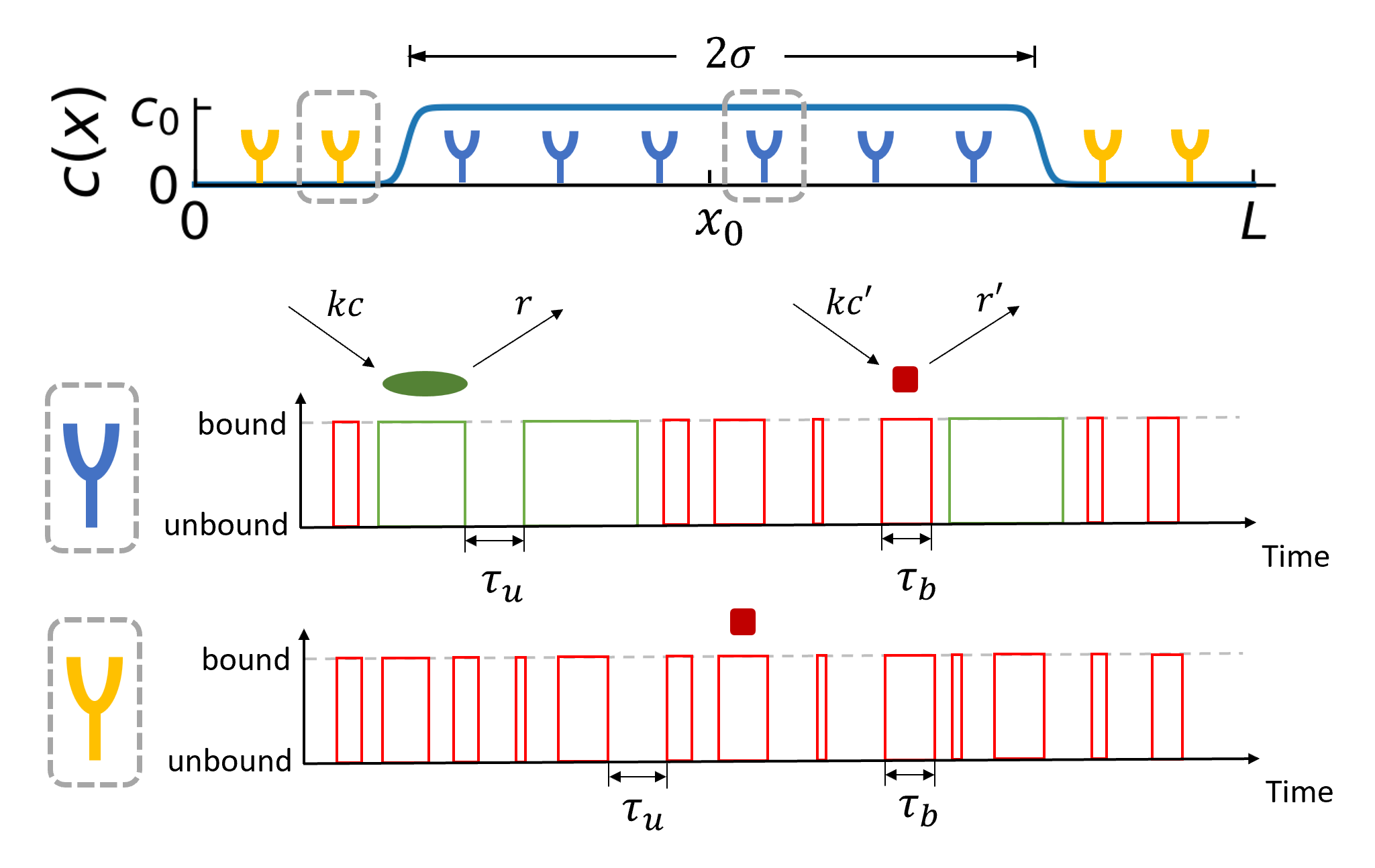}%
\caption{\label{fig:time_series} Extracting information from bound-unbound time series of receptors. Top: The ephrin concentration is a smooth rectangular function $c(x)$ that is high at the region of contact ($\sim \pm \sigma$ of the contact center at $x_0$). Bottom: The temporal record of the binding states of example receptors inside and outside the contact region.  Receptors within the contact region (blue) can bind to both ephrins (green) and incorrect ligands (red) while receptors outside (yellow) can only bind to the spurious ligands. }
\end{figure}

{As an initial model for the fundamental sensing limits of contact inhibition of locomotion, we work in an effectively one-dimensional model, with Eph receptors {characterized by their location $x$} around the perimeter of cell 1.}
We assume ephrins {from cell 2} can bind to the Eph receptors {on cell 1} with a rate $kc(x)$, i.e., a base rate $k$ times an effective ephrin concentration $c(x)$. {The rate $kc(x)$ combines both the area density of the ephrin on the membrane with the reduced likelihood of binding for ephrins which are further away from the membrane of cell 1 -- {so if we treat $k$ as constant} we expect $c(x)$ to vary from its maximum value at the cell-cell contact to essentially zero far away from the contact. The details of the shape of $c(x)$ will depend on the details of the cell-cell junction and the cell-cell contact, but we make an initial hypothesis that $c(x)$ is effectively constant over the cell-cell contact width and then smoothly decreases to zero (Fig. \ref{fig:time_series}). We use a smooth rectangular function for the effective ephrin concentration that cell 1 senses on cell $2$'s surface}:
\begin{equation}\label{eq:cx}
    c(x)=c_0 S(x-(x_0-\sigma)) S((x_0+\sigma)-x),
\end{equation}
where $S(x)=[1+\tanh(x/\omega)]/2$, and $\omega$ controls the steepness at the transition between $0$ and $c_0$.
$x_0$ is the center of the region of cell-cell contact and $2\sigma$ is the contact width. {Receptors bound with the correct ligand (ephrin) will unbind with rate $r$.}
{Receptors can also bind the incorrect/spurious ligand with rate $kc'$ and unbind with rate $r'$ (Fig. \ref{fig:time_series}), where we assume the effective concentration of incorrect ligands $c'$ is constant.}

{In our model,} we adopt the assumption of Ref. \cite{mora}, where the only distinction between the two ligands is their binding kinetics, neglecting features like potential ligand bias \cite{karl2020ligand, gomez2022ligands}. {This means that the cell only knows whether a receptor is bound, not what ligand is bound to it. How, then, can the cell detect the presence of another cell and repolarize away from it? First, even if ligands had identical unbinding rates, there would be more bound receptors at the cell-cell contact. Secondly, if the correct and incorrect ligands have different affinities ($r \neq r'$), the cell can discriminate between them statistically. } 
Ligands with a larger unbinding rate remain bound to receptors for a shorter time compared to ligands with a smaller unbinding rate. Thus, the cell can discern the mixture of two different ligands if the detailed occupancy history of each receptor is available \cite{mora}. {This record can be summarized by the times the receptor spends unbound $\tau_{u,i}$ and the times it spends bound $\tau_{b,i}$ \cite{PhysRevLett.104.248101,mora,endres2009maximum}, e.g., $\{\tau_{u,1},\tau_{b,1},\tau_{u,2},\tau_{b,2},\cdots\}$} (as shown in Fig. \ref{fig:time_series}). Given the receptor record, $c(x)$ can be estimated {which allows the cell to identify the location of the cell-cell contact $x_0$.} 

The probability densities to have a bound interval of length $\tau_b$ and an unbound interval of length $\tau_u$ for a receptor seeing concentration $c$ of correct ligand are~\cite{mora, PhysRevResearch.2.043146}
\begin{eqnarray}
    &&P(\tau_u)=k(c+c')\ee^{-k(c+c')\tau_u},\label{eq:p_tu}\\
    &&P(\tau_b)=\frac{c}{c+c'}r\ee^{-r\tau_b} + \frac{c'}{c+c'}r'\ee^{-r'\tau_b},\label{eq:p_tb}
\end{eqnarray}
where the factor $(c+c')$ in Eq. \eqref{eq:p_tu} is the total concentration, and the factors $c/(c+c')$ and $c'/(c+c')$ in Eq. \eqref{eq:p_tb} {are the probabilities of the binding event being the correct ligand and incorrect ligand, respectively}. We assume that there are $N$ receptors {evenly spaced over} the cell perimeter $L$, {each with position $x_n$ 
so the $n$-th receptor has ephrin concentration 
$c(x_n)$.} 

{We assume that the cells observe the binding and unbinding events over a ``measuring time'' $T$ and use this to make a CIL decision -- estimating the position of the cell-cell contact. The probability of a particular time record $\{\tau_{u,1},\tau_{b,1},\tau_{u,2},\tau_{b,2},\cdots\}$ for the $n$-th receptor is $P_n=\prod_{i=1}^{M_n}P(\tau_{u,i}^{(n)})\cdot P(\tau_{b,i}^{(n)})$ where $M_n$ is the number of pairs of binding-unbinding events during the time $T$}. Therefore, given the three parameters $\boldsymbol{\theta}\equiv(c_0, c', x_0)$ in our model, the total probability of a trajectory of bound and unbound times arising from $N$ independent receptors is $P(\{\tau\}|\boldsymbol{\theta})=\prod_{n=1}^N P_n$.

To derive the fundamental sensing limits, we use the method of maximum likelihood estimation (MLE) \cite{endres2009maximum,PhysRevE.83.021917,kay1993fundamentals}.
The log-likelihood function is $\ln\mathcal{L}(\boldsymbol{\theta}|\{\tau\})=\ln P(\{\tau\}|\boldsymbol{\theta})$. The MLE estimate of the parameters {$\hat{\boldsymbol{\theta}}$ is obtained by maximizing the log-likelihood, e.g., by solving $\partial\ln\mathcal{L}/\partial \theta_\mu|_{\boldsymbol{\theta} = \hat{\boldsymbol{\theta}}}=0$, where the index $\mu=1,2,3$, or by numerical maximization of $\ln\mathcal{L}$}.
{The error between any unbiased estimate of the parameters that the cell can make, $\hat{\boldsymbol{\theta}}$, and the true parameters is limited by the Fisher information matrix and the Cram\'er-Rao bound (Eq. \ref{eq:CRB}). We can compute the Fisher information matrix $I(\boldsymbol{\theta})_{\mu\nu}=-\langle\partial^2\ln\mathcal{L}/\partial\theta_\mu\partial\theta_\nu\rangle$ for our model in a partially-analytic way. After taking the derivatives of the log-likelihood necessary to compute the Fisher information matrix, we get an equation that is a complicated function of the bound/unbound times $\{\tau_b,\tau_u\}$. This reduces the problem to quadrature -- we can calculate the average $\langle\cdots\rangle$ by numerical integration using the probability densities in Eq. \eqref{eq:p_tu} and \eqref{eq:p_tb} (see Appendix \ref{app:fisher} for details). The  Cram\'er-Rao bound (CRB) then gives the lower bound for the variance of an unbiased estimator  $\hat{\boldsymbol{\theta}}$ of parameters~\cite{PhysRevE.105.044410}:}
\begin{equation}\label{eq:CRB}
    \textrm{Var}({\hat{{\theta}}}_\mu)\geqslant [I(\boldsymbol{\theta})^{-1}]_{\mu\mu},
\end{equation}
{where $I(\boldsymbol{\theta})^{-1}$ is the matrix inverse.}
{Eq. \eqref{eq:CRB} sets the best precision with which a cell could measure, e.g., the location of the cell-cell contact, given the unavoidable stochasticity arising from the ligand-receptor interactions.}

\section{Results}

{How precisely can a cell detect the position of cell-cell contact, or the concentrations of ligands? This is set by Eq. \eqref{eq:CRB} -- the best possible unbiased estimates we can make of our parameters. We will display our results in terms of the variables} %
$\boldsymbol{\theta}'\equiv(c_t,\chi,x_0)$, {where $c_t\equiv c_0+c'$ is the total concentration, i.e., the concentration that sets the total binding rate at the cell-cell contact and $\chi=c_0/c_t$ is the fraction of correct ligands. When $\chi \ll 1$, most bindings are the incorrect ligand, and we expect the cell to struggle to perform CIL.}
We also define the unbinding rate ratio of the two ligands as $\alpha\equiv r/r'$. The correct ligands are expected to have a higher affinity to the receptors and thus take longer to unbind, i.e., $r\leqslant r'$, so we assume $\alpha\leqslant 1$. 

\begin{figure*}
\includegraphics[width=0.8\textwidth]{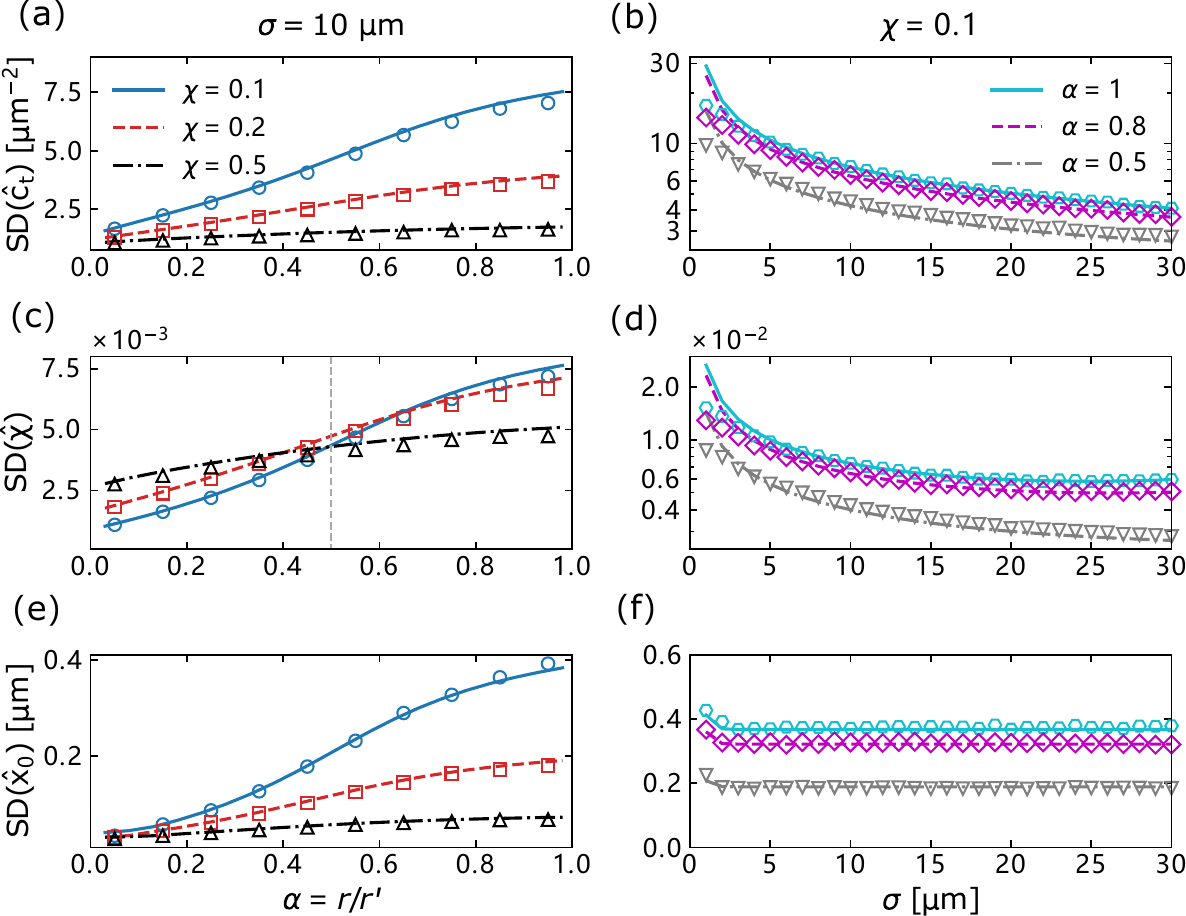}%
\caption{\label{fig:var} Errors of maximum likelihood estimate (MLE) influenced by receptor-ligand affinity, fraction of correct ligands, and contact width. (a-b) show the standard deviation (SD) of the total concentration $\hat{c}_t$, (c-d) show the SD of correct ligand fraction $\hat{\chi}$, and (e-f) display the SD of contact center position $\hat{x}_0$. In all figures, the lines (solid, dashed, and dash-dot) represent theoretical results obtained from  Eq. \eqref{eq:CRB}, while the points are obtained from Monte Carlo simulations (the standard deviations are computed from $10^4$ simulations for each point); Measuring time is $T=10\,\rm{s}$.}
\end{figure*}

In Fig. \ref{fig:var}, we plot both our calculations of the minimal standard deviations required from the Cram\'er-Rao bound in Eq. \eqref{eq:CRB} as well as simulation results from Monte Carlo (MC) simulations {where we generate stochastic receptor trajectories according to Eq. \eqref{eq:p_tu} and Eq. \eqref{eq:p_tb}, and then numerically maximize the likelihood to determine the maximum-likelihood estimators $\hat{c_t}$, $\hat{\chi}$, and $\hat{x}_0$ (Appendix \ref{app:stochastic})}. Broadly, we see good agreement between simulation and theory, except for a few data points at small patch sizes when $\sigma \sim 1-2\,\rm{\mu m}$, which we discuss further below. 

\subsection{How do the sensing limits change when it becomes difficult to distinguish the two ligands?}

The unbinding rate ratio $\alpha=r/r'$ quantifies the difference in receptor-ligand affinity between the two ligands. In other words, it characterizes how hard it is for the cell to discriminate the two ligands. When $\alpha$ is close to $0$, i.e., in the easy discrimination regime, the standard deviations of all three estimated parameters $(\hat{c}_t,\hat{\chi},\hat{x}_0)$ are small. When $\alpha$ gets closer to $1$, it becomes harder to distinguish the two ligands. If $\alpha=1$, both ligands have the same unbinding rate and are thus indistinguishable for the cell within our assumptions. The difficulty to tell the two ligands apart leads to a rise in sensing errors for all parameters as $\alpha$ nears 1, i.e., the smaller the discrepancy between ligands, the less accurately the cell can sense the contact (Fig. \ref{fig:var}). 
The standard deviations of all three parameters are small compared to their {relevant scales, which are for Fig. \ref{fig:var} total concentration $c_t=1000\,\rm{\mu m^{-2}}$, correct ligand fraction $\chi=0.1$, and the cell perimeter $L=100\,\rm{\mu m}$}. {This means that the limits imposed by stochastic noise of ligand-receptor binding may not be that stringent in practice.} For example, if the cell can locate the source of cell-cell contact to $\pm \textrm{SD}(\hat{x}_0)$ of $\sim0.4$ microns (the worst case of Fig. \ref{fig:var}f), this means the cell has an angular uncertainty of the direction to go which is $2\pi\cdot\textrm{SD}(\hat{x}_0)/L$ of $\sim 1.5^\circ$. This angular uncertainty is much smaller than the typical angular range of acceleration vectors observed in CIL {on two-dimensional substrates}~\cite{exp}, which indicates that cells likely have more positional information than they really use in CIL -- unless the concentration of spurious ligands is extraordinarily high ($\chi \ll 1$) or {our model is incomplete or there is additional noise due to }one of the other factors we raise later in Sec. \ref{sec:discussion}. 

In Fig. \ref{fig:var} (a,c,e), we also show the standard deviations for different correct ligand fractions $\chi=c_0/c_t$. {We expect that, as the fraction of correct ligands decreases, the error in estimating each parameter should increase.} We do see that errors associated with estimating the total concentration $c_t$ and the contact center position $x_0$ decrease as $\chi$ increases (Fig. \ref{fig:var}a,e). {However, estimating the correct ligand fraction $\chi$ itself has a more complex behavior (Fig. \ref{fig:var}c). Estimating $\chi$ in our model is akin to a spatially-dependent multi-receptor model of the question asked by \cite{mora} -- how precisely can a single receptor detect a rare ligand binding? Our model, though, shows qualitative differences compared to the single receptor model of \cite{mora}. For a single receptor in the easy discrimination regime when the cell is confident about what type of ligand is bound ($\alpha<1/2$), the uncertainty of sensing $\chi$ increases with $\chi$ as $\textrm{SD}(\hat{\chi}) \sim \chi^{\beta/2}$ with $\beta = 1-\alpha/(1-\alpha)$, though the relative error $\textrm{SD}(\hat{\chi})/\chi$ always decreases with $\chi$ \cite{mora}, while in the hard discrimination regime ($\alpha > 1/2$), $\textrm{SD}(\hat{\chi})$ is roughly independent of $\chi$, and diverges as $\alpha \to 1$, where a single receptor cannot discriminate between receptors. In our spatial, multi-receptor model, we find that when $\alpha < 1/2$ the error in $\chi$ increases with $\chi$, as seen in \cite{mora}.
However, in the hard discrimination regime ($\alpha>1/2$), we find that $\textrm{SD}(\hat{\chi})$ has the opposite behavior, increasing when $\chi$ decreases. We argue that this occurs because there are two ways for the cell to estimate $\chi$: first, by discriminating between the two ligands by their binding times, and secondly, by comparing the total amount of binding observed at the cell-cell contact (depending on $c_0 + c')$ to the binding outside the contact (depending only on $c'$). The second mechanism does not require that the cell can discriminate between the two ligands, and will still function as $\alpha \to 1$. Hence, we see that as $\alpha \to 1$, the error of sensing $\chi$ decreases as $\chi$ increases just like the error of the total concentration $c_t$.} This source of error is essentially arising because at small $\chi$, the difference between $c_0 + c'$ and $c_0$ will shrink. By contrast, for $\alpha < 1/2$, the estimation of $\chi$ is similar to the single-molecule discrimination of \cite{mora} and the error in $\chi$ increases with $\chi$.

\subsection{How does contact width limit CIL?}
{As the contact width $\sigma$ expands, more receptors come inside the contact region and the cell can obtain more information about the concentration of the true ligands -- we expect that the accuracy of estimating our parameters should generally increase as the contact width increases. }
In Fig. \ref{fig:var} (b) and (d), we observe that both the standard deviation of total concentration $\textrm{SD}(\hat{c}_t)$ and the standard deviation of the fraction of correct ligands $\textrm{SD}(\hat{\chi})$ decrease with increasing contact width $\sigma$.
{Surprisingly,} we find the error in estimating the contact position $\textrm{SD}(\hat{x}_0)$ remains almost constant as $\sigma$ ranges from 3-30 $\mu$m (Fig. \ref{fig:var}f). {Why? We argue that the cell's information about the contact location is largely coming from where the edges of the contact are, i.e., the transitions in $c(x)$.} To better illustrate this, consider a simpler model that only involves one snapshot in time and one species of ligands, {which we denote with a capital $C(x)$. In this toy model, given concentration} $C(x)$, the occupancy for each receptor is given by $p_n=C(x_n)/[C(x_n)+K_D]$, with the dissociation constant $K_D=r/k$~\cite{PhysRevLett.105.048104}. The joint probability for the state of all receptors is $P(\{Z_n\}) = \mathcal{L}^\textrm{toy}$ with
\begin{equation}
    \mathcal{L}^\textrm{toy} =\prod_n p_n^{Z_n}(1-p_n)^{1-Z_n},
\end{equation}
where $Z_n = 0$ indicates the receptor is unbound and $Z_n = 1$ indicates it is bound.
In this toy model, we can compute the Fisher information:
\begin{equation}
\begin{split}
        I^\textrm{toy}_{\mu\nu}&=-\left< \frac{\partial^2\ln \mathcal{L}^\textrm{toy}}{\partial\theta_\mu\partial \theta_\nu} \right> \\
    &=\sum_{n}\frac{K_D}{[C(x_n)+K_D]^2 C}\frac{\partial C(x_n)}{\partial \theta_\mu}\frac{\partial C(x_n)}{\partial \theta_\nu}. \label{eq:I_snapshot}
\end{split}
\end{equation}
{If the concentration $C(x)$ takes the form of a smooth rectangular function (Eq. \ref{eq:cx}), and we compute the $x_0$-$x_0$ component of the Fisher information matrix, the only terms that are nonzero in the sum in Eq. \eqref{eq:I_snapshot} are those where the derivative $\partial C(x) / \partial x_0$ is nonzero -- which are the edges of the contact zone. By contrast, if we were going to estimate the concentration level of the contact, then you'd get a non-zero term at every receptor within the contact, and we'd expect the error to decrease with contact size.}

{In Fig. \ref{fig:var} (f) the uncertainty in measuring the contact center $x_0$ does increase for small enough $\sigma$, in the range of $\sigma = 1-2 \,\rm{\mu m}$. %
This happens when the contact width $\sigma$ is similar to the scale $\omega = 1\,\rm{\mu m}$ over which $c(x)$ is varying -- so then $\partial c/\partial x_0$ will be nonzero over the whole contact patch.}

{For most of Fig. \ref{fig:var}, we see good agreement between our bound from the Fisher information in Eq. \eqref{eq:CRB} and the variance of the maximum likelihood estimator computed from a stochastic simulation. However, at the smallest $\sigma$, we do see a larger deviation between Monte Carlo simulations and Fisher information results in Fig. \ref{fig:var} (b,d,f). This discrepancy likely arises from the difficulty of the maximization process when $\sigma$ is small -- i.e., when we do numerical optimization to find the maximum-likelihood estimator $\hat{\boldsymbol{\theta}}$, the optimization converges to a local maximum instead of finding the correct maximum likelihood estimator. These numerical problems are more likely to happen for small contact size $\sigma$ and $\alpha \to 1$, because in such cases, there is less difference between a cell with a very small contact patch and one with essentially no contact ($\chi = 0$). In fact, the reasons that make it difficult for us to do this numerical optimization also make it difficult for the cell to determine if another cell is present, which we address in the next section.}

\subsection{How reliably can the cell detect contact at all?}

In many situations, determining the presence of another cell is more useful for the cell than precisely locating the contact point. For instance, a recent experiment \cite{singh2021rules} studied CIL for cells on suspended nanofibers, which are used to mimic the extracellular matrix. In this biologically relevant case -- and in other experiments of CIL on micropatterns \cite{desai2013contact,lin_interplay_2015} -- cells only have two possible directions. Thus the precision in estimating the location of contact $x_0$ is not so important, but deciding whether there is contact or not is very important.

We study the problem of how precisely a cell can detect the presence of a neighbor from a localized signal by using tools from statistical decision theory.
Firstly, if we regard the cell as a binary classifier, we can pose a question to the cell: ``Is there another cell  in proximity?'' This question can be formally translated into the task of distinguishing between two hypotheses: the presence vs. absence of another cell. In our model, if we denote the three-dimensional parameter space where $\boldsymbol{\theta}$ lives as $\Theta$, the two hypotheses -- presence or absence~\cite{mora, likelihoodRatio, lalanne2015chemodetection}, can be written as:
\begin{eqnarray}
&&H_0: \boldsymbol{\theta}\in\Theta_0\equiv\{(c_0,c',x_0)|c_0=0\},\\
&&H_1: \boldsymbol{\theta}\in\Theta_0^C\equiv\Theta\backslash\Theta_0,
\end{eqnarray}
where $H_0$ (called the null hypothesis) states $\boldsymbol{\theta}$ lives within a subspace $\Theta_0$, which corresponds to the plane of $c_0=0$, i.e., the other cell is not there. And $H_1$ (called the alternative hypothesis) asserts $\boldsymbol{\theta}$ lives in $\Theta_0^C$, the complement space of $\Theta_0$, i.e., $c_0\neq 0$ and there is another cell nearby. 

\begin{figure*}
\includegraphics[width=0.8\textwidth]{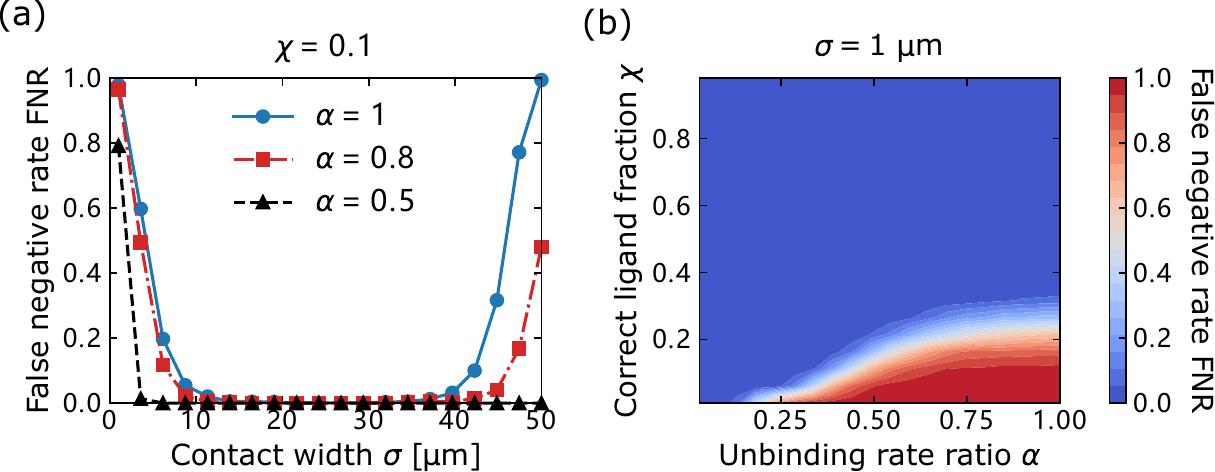}%
\caption{\label{fig:FNR} False negative rates showing how the probability of a cell {missing the presence of a contacting cell} is affected by contact width $\sigma$, unbinding rate ratio $\alpha$, and the fraction of correct ligands $\chi$. (a) illustrates the change in FNR as contact width varies, and the phase diagram (b) shows a sharp transition from low $\alpha$, high $\chi$ values to high $\alpha$, low $\chi$ values. We conduct $\sim 10^5$ simulations for each data point. The threshold parameter is set to $\lambda_c=7$ is chosen somewhat arbitrarily to show when the maximum value of FNR can be close to $1$ in these figures.}
\end{figure*}

This question can be addressed by employing a statistical method known as the likelihood-ratio test (Wilks test)~\cite{likelihoodRatio} to help the cell make decisions. The basic concept of this test is to calculate the ratio of the maximum likelihood under the two competing hypotheses:
\begin{eqnarray} \label{eq:lambda}
    \lambda = \ln\frac{\sup_{\boldsymbol{\theta}\in\Theta}\mathcal{L}}{\sup_{\boldsymbol{\theta}\in\Theta_0}\mathcal{L}}=\sup_{\boldsymbol{\theta}\in\Theta}(\ln\mathcal{L})-\sup_{\boldsymbol{\theta}\in\Theta_0}(\ln\mathcal{L})\geqslant0,
\end{eqnarray}
where the $\sup$ notation refers to the supremum. Therefore, if $\lambda$ is large, it indicates the alternative hypothesis $H_1$ (presence) is more competitive. In contrast, the null hypothesis $H_0$ (absence) becomes a more compelling choice if $\lambda$ is small. 
In general, the ratio $\lambda$ will tune how much evidence is needed to prefer hypothesis $H_1$ over hypothesis $H_0$.
Then we can set an adjustable threshold value $\lambda_c$ for the cell. If $\lambda>\lambda_c$, the cell {believes it is contacting another cell. Conversely, if $\lambda < \lambda_c$}, the cell concludes that there is no other cell nearby.

To gain deeper insight into the problem, we want to know: How likely is it for the cell to fail in detecting the presence of another cell when it is indeed there? Or how likely is it for the cell to claim the detection of another cell but it is actually absent? In statistics, the former probability is defined as the false negative rate (FNR), and the latter is referred to as the false positive rate (FPR). In our model, the false negative rate can be computed as~\cite{mora}
\begin{equation}\label{eq:FNR}
\textrm{FNR}=\langle H(\lambda_c-\lambda)\rangle_+,
\end{equation}
where $H(x)$ is the Heaviside step function. The symbol $\langle\cdots\rangle_+$ denotes the average over all allowed configurations of the time series, and the positive subscript $+$ indicates the average is taken under the condition of another cell being there. Only when the cell misses the presence of the other cell, i.e., $\lambda<\lambda_c$, the step function is non-zero. Thus, Eq. \eqref{eq:FNR} gives exactly the probability of the cell making a false negative decision, where it erroneously asserts the absence of the other cell. {To compute the FNR, we choose a value $\lambda_c$, use our Monte Carlo simulation to generate a set of receptor occupancies over time given the presence of a second cell (Appendix \ref{app:stochastic}), then numerically maximize the likelihood under the two hypotheses (presence and absence of the contacting cell) to compute $\lambda$ from Eq. \eqref{eq:lambda}. We then repeat this many times to compute the FNR from Eq. \eqref{eq:FNR}.} 

Fig. \ref{fig:FNR} (a) shows the Monte Carlo simulation results of FNR for varying contact widths. 
{At small $\sigma$}, when there are relatively few receptors where the true concentration $c(x)$ is nonzero,
the false negative rate is large. Essentially, in this case, the probability of small numbers of correct ligands binding from cell-cell contact being mistaken for a coincidence is large. The FNR decreases as the contact width is increased, leading more receptors to come into the contact region,
However, when the contact width $2\sigma$ becomes very large and approaches the cell perimeter $L$, the false negative rate starts to rise again, particularly when $\alpha=1$. This $2\sigma\to L$ case could happen if the cell is nearly engulfed by another cell, or is in contact with many cells.
{FNR will increase as the contact region approaches $L$ only} in the hard discrimination regime ($\alpha$ approaches 1), when the two ligands are indistinguishable. {When $2\sigma$ approaches the cell perimeter, $c(x)$ becomes nearly constant $c(x) \approx c_0$ -- equivalent to a simple change in the background concentration $c'$, if the cell cannot discriminate between the two ligands. The rise in FNR at large $\sigma$ disappears as the cell becomes increasingly capable of distinguishing between the two ligands, i.e., when $\alpha<1$.}

\begin{figure*}
\includegraphics[width=0.85\textwidth]{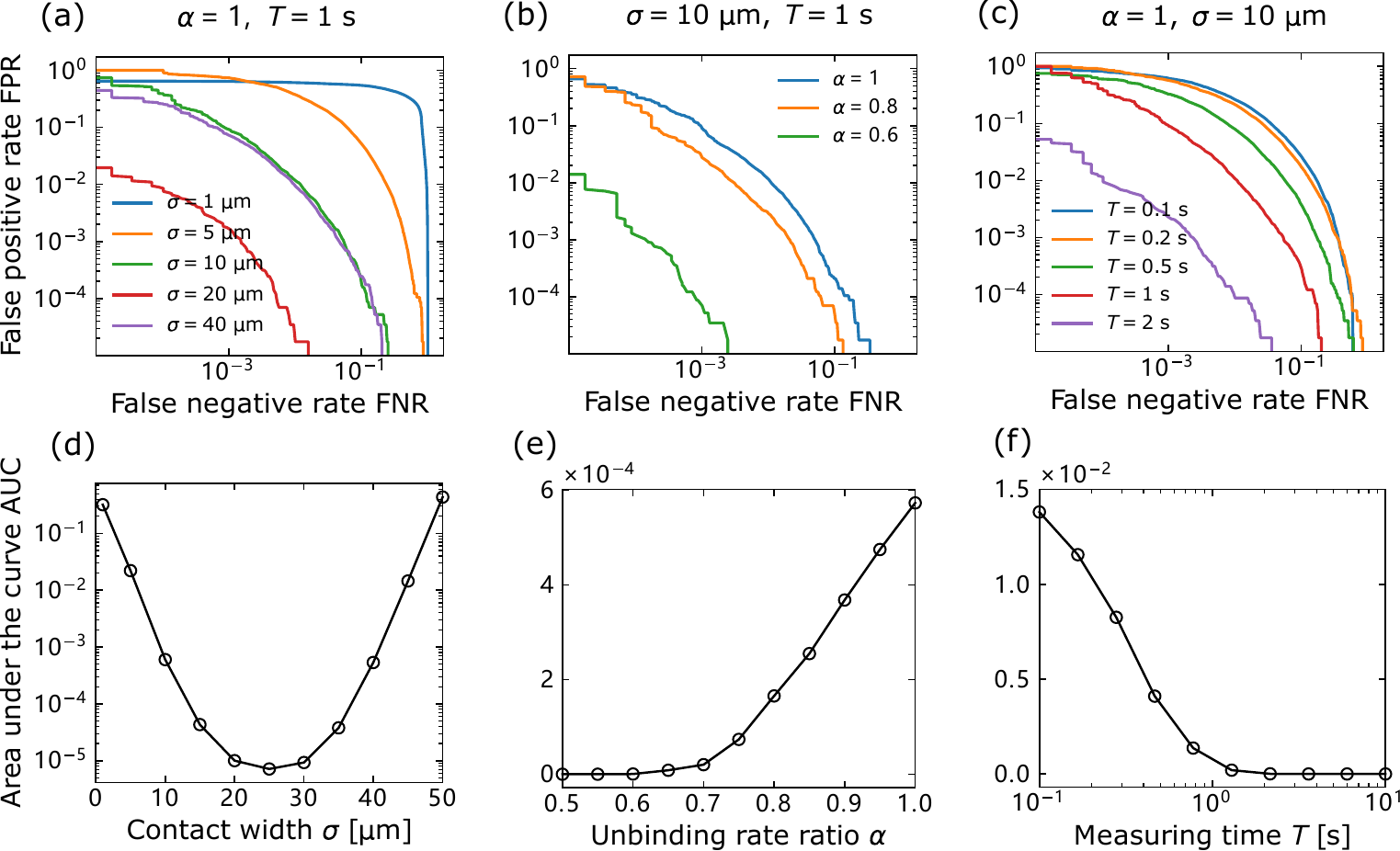}%
\caption{\label{fig:ROC} Detection error tradeoff (DET) graphs and the associated area under the curve (AUC) showing how the cell's ability of making accurate decisions varies with the contact width $\sigma$, unbinding rate ratio $\alpha$, and measuring time $T$. (a) and (d) show that the optimal contact width is at some intermediate value. (b) and (e) show that the cell is more likely to make incorrect decisions when it's difficult to distinguish the two ligands. (c) and (f) show that the cell can more reliably detect the presence or absence of another cell with a longer measuring time $T$.
Correct ligand fraction is set to $\chi=0.1$ in all figures.}
\end{figure*}

We then perform a sweep across the unbinding rate ratio $\alpha$ and the correct ligand fraction $\chi$, both of which significantly impact the false negative rate. We show a phase diagram for when cells might fail to recognize a contact in Fig. \ref{fig:FNR} (b), which shows a sharp transition in FNR as we go from low $\alpha$, high $\chi$ values to high $\alpha$, low $\chi$ values. Notably, the FNR exhibits non-zero values only when $\alpha$ approaches $1$ and $\chi \ll 1$, i.e., the cell is more prone to making mistakes in the hard discrimination regime, where the ligands are nearly indistinguishable, and when there is a higher presence of interfering background ligands. These findings align with the previous results in Fig. \ref{fig:var}.

The false negative rates are influenced by the adjustable threshold parameter $\lambda_c$ we chose. We can always increase the false positive rate (FPR) and decrease the false negative rate by adjusting the threshold parameter $\lambda_c$ to a small value, and vice versa. To compare the ability to discriminate independent of the arbitrary $\lambda_c$, we use the detection error tradeoff (DET) graph~\cite{DET, ROC, mora}.  First, we can compute the false positive rate in a similar manner to $\textrm{FNR}$:
\begin{equation}\label{eq:FPR}
\textrm{FPR}=\langle H(\lambda-\lambda_c)\rangle_{-},
\end{equation}
where $\langle\cdots\rangle_{-}$ indicates taking the average {with no second cell present, i.e., setting the ephrin concentration $c_0 = 0$.}
We visualize the trade-off between false positives and false negatives by varying the threshold parameter $\lambda_c$ over {the whole range of $\lambda$ observed in our simulations}, and then plotting the FPR found for each $\lambda_c$ against the FNR, as shown in Fig. \ref{fig:ROC} (a-c). 
{The figures in Fig. \ref{fig:ROC} (a-c) show how FPR varies as a function of FNR. For instance, if we look at Fig. \ref{fig:ROC} (a), the contours indicate the best false positive rate the cell can achieve given a fixed false negative rate, or vice versa. For instance, if the cell can tolerate a false positive rate of $10^{-3}$, and the contact width is $\sigma=20\,\mathrm{\mu m}$, the cell can achieve a low false negative rate (down to $\sim 10^{-3}$, red line). However, either decreasing the contact to $\sigma=10\,\mathrm{\mu m}$, or expanding the contact to $\sigma=40\,\mathrm{\mu m}$ makes the FNR larger for this FPR -- the cell is more likely to overlook the contact.}

{In Fig. \ref{fig:ROC} (b), we see that the ability of the cell to reliably distinguish the presence of contact gets progressively worse as $\alpha \to 1$ -- the FPR-FNR tradeoff curve gets further from the origin. Increasing the cell's measurement time $T$ improves the accuracy with which the cell can sense cell-cell contact (Fig. \ref{fig:ROC}c).}

{We can summarize the efficacy of the cell's ability to distinguish the presence of a contacting cell by computing the area under the curve (AUC) of the tradeoff graphs in Fig. \ref{fig:ROC} (a-c)}. A smaller AUC indicates that the cell exhibits higher accuracy in distinguishing between the two hypotheses $H_0$ and $H_1$ -- {roughly, that the cell achieves a lower false negative rate for a given false positive rate}. {We note that tradeoffs of this sort are also often plotted on a receiver operating characteristic (ROC) curve, which plots true positive rate against false positive rate. In our convention, a smaller area under the curve is more predictive, instead of larger; our AUC is $1-\textrm{AUC}_{\textrm{ROC}}$}~\cite{DET, ROC}.

Fig. \ref{fig:ROC} (d) shows the AUC value as a function of contact width, which initially decreases and then increases as $2\sigma$ gets closer to $L$. This confirms the results of Fig. \ref{fig:FNR} (a): the cell can better sense the existence or absence of the other cell when the contact width is at some intermediate value. Fig. \ref{fig:ROC} (b) and (e) show how the DET curve and the associated AUC change with the unbinding rate ratio $\alpha$. The area under the curve increases as $\alpha\to 1$, which means the cell is more likely to make incorrect decisions in the hard discrimination regime. We also observe that the AUC drops to approximately $0$ in the easy discrimination regime when $\alpha\approx 0.5$, which is consistent with where the transition happens in the phase diagram of FNR (Fig. \ref{fig:FNR}b). {As we saw above, accuracy in sensing also depends on the fraction of ligands that are spurious. If we decrease the correct ligand fraction to $\chi=0.01$ (Fig. \ref{fig:chi_AUC}), the overall AUC can become much larger, reaching $\textrm{AUC} \sim 0.4$, i.e., it becomes much harder to make correct decisions. However, even with this extremely small correct ligand fraction $\chi$, the cell can detect contact very accurately once the unbinding rate ratio reaches $\alpha\sim0.25$ (Fig. \ref{fig:chi_AUC}). So far, we have kept the amount of time that the cell uses to make a decision constant. Increasing the measurement time $T$ monotonically brings the DET curve toward the origin, decreasing the AUC 
(Fig. \ref{fig:ROC}c,f). Cells can more reliably detect the presence or absence of another cell by extending the measuring time $T$, as we would expect. AUC are mostly large when the measuring time is on the scale of seconds or less. This is a relatively short time compared to the time over which CIL takes place, which can be tens of minutes~\cite{desai2013contact}.}

\section{Discussion}\label{sec:discussion}

Our multi-ligand model highlights the significant impact of various factors on the accuracy of sensing cell-cell contact, including the receptor-ligand affinity, the fraction of correct ligands, and the contact width. It becomes more challenging to sense the cell-cell contact when it's difficult for the cell to distinguish ephrins from spurious ligands, or when there are more interfering spurious ligands present. Moreover, our model reveals that as the contact width $\sigma$ expands, it becomes easier for the cell to estimate the concentration and the fraction of correct ligands. However, the error in contact position $x_0$ estimation remains constant while the contact width increases. We argue that this phenomenon occurs because the cell can only obtain spatial information from the two boundaries where the concentration changes.
Furthermore, we investigate whether the cell can successfully detect the presence of another cell through a small cell-cell contact. Our finding indicates that the cell is more likely to make incorrect decisions when the contact width is either too small or too large. In addition, our results also suggest that the cell is more probable to make mistakes if it's difficult for the cell to distinguish ephrins from spurious ligands ($\alpha \to 1$). The cell can extend the time it takes to make a decision to improve decision-making robustness.

{A core element of our model is the profile $c(x)$ in Eq. \eqref{eq:cx}, which represents a cell in contact with another cell with a clear contact region of fixed size and a smoothed rectangular form. This neglects many potential complications -- e.g., that the size of cell-cell contacts may evolve over time during CIL (see, e.g., the movies of \cite{singh2021rules}), and how ligand-receptor binding could drive changes in the size and spacing of the cell-cell contact \cite{hu2013binding,dinapoli2021mesoscale}. To check the robustness of some of our assumptions, we also explore the case where the ephrin concentration takes the form of a Gaussian function $c(x)=c_0\exp(-(x-x_0)^2/2\sigma^2)$. This would arise if we assume that our effective binding rate $kc(x)$ is suppressed when there is a gap large enough to require ephrin to undergo a strain in order to bind. If there is a gap $h(x)$ between the two cells, we'd expect $kc \sim \ee^{-\frac{1}{2}\kappa h^2 / k_B T}$ with $\kappa$ an effective spring stiffness \cite{janevs2022first}. If  $h(x)$ increases linearly away from the point of closest approach $x_0$, we would get a Gaussian profile.}
{Many of our core results are quite similar between the Gaussian and the smooth step profile for $c(x)$, see Fig. \ref{fig:Gaussian} (a-e).
However, in the scenario of a Gaussian concentration, we find that the error in contact position estimation $\textrm{SD}(\hat{x}_0)$ increases with increasing contact width (Fig. \ref{fig:Gaussian}f).}
Unlike the case of rectangular profiles, the cell now can obtain spatial information from receptors away from the contact edges, as $\partial c/\partial x_0=c(x)(x-x_0)/\sigma^2$ is nonzero throughout the contact and depends systematically on $\sigma$ -- {there is more spatial information at any given point as the width $\sigma$ becomes smaller}. Therefore, the error of estimating the contact position $x_0$ doesn't remain constant as $\sigma$ expands; instead, it increases with increasing $\sigma$.

The estimation errors shown in our model (Fig. \ref{fig:var}) {are generally small compared to their relevant scales, suggesting cells may have fairly precise knowledge of the cell-cell contact location and ephrin levels. This may reflect the large number of binding-unbinding events. With the total number of receptors $N=10^4$ and typical receptor correlation times of $\sim \mathrm{s}$ (Appendix \ref{app:correlation_time}), and the measuring time $T=10\,\mathrm{s}$, there are $\sim 10^5$ binding-unbinding events during the time the cell makes its CIL measurement.} {Given our results, it is likely that, unless one of our assumptions is incorrect, or that the fraction of spurious ligands is extremely high (e.g., $\chi < 0.1$), stochastic ligand-receptor noise is not a crucial factor in CIL decisions.}  {Where could our assumptions be incomplete?} One possibility is that we have neglected details of the Eph receptor interactions. Experiments have found that receptor clustering and other biophysical complexities~\cite{zapata-mercado_efficacy_2022, doi:10.1073/pnas.0707080105, singh_epha2_2018, PhysRevLett.107.178101} can influence the functioning of receptors in binding kinetics in Eph receptors and other RTKs. For instance, EGF receptors have been observed to dimerize while binding to their ligands EGF~\cite{doi:10.1073/pnas.0707080105}. Moreover, for A549 lung cancer cells, it has been observed that the vast majority of EphA2 molecules exist in clusters~\cite{singh_epha2_2018}. As a result, receptor clustering could significantly reduce the number of distinct receptor locations in our model, thereby substantially increasing the errors when estimating the cell-cell contact.
Additionally, the precise values of the binding constant $k$ and unbinding rate $r$ for ligands and receptors tethered to cell membranes, such as Eph-ephrin, are not entirely clear~\cite{bihr_association_2014, PhysRevX.12.031030, zarnitsyna_t_2012, bicknell_limits_2015}. These parameters can also significantly influence the number of binding-unbinding events $M$ that occur during the measuring time $T$.
 {Sources of noise downstream of the initial ephrin binding, such as stochasticity in the polarization of the cell's Rho GTPases \cite{walther2012deterministic,kulawiak2016modeling,copos2020hybrid,ipina2023secreted}, might lead to significant additional noise in responses to CIL.} Incorporating these factors into our model could provide a more comprehensive understanding of cell-cell interaction and response mechanisms.

\begin{acknowledgments}
The authors acknowledge support from NIH grant R35GM142847. This work was carried out at the Advanced Research Computing at Hopkins (ARCH) core facility  (rockfish.jhu.edu), which is supported by the National Science Foundation (NSF) grant number OAC 1920103. We thank Pedrom Zadeh and Grace Luettgen for a close reading of the manuscript. 
\end{acknowledgments}

\appendix

\section{Correlation time of binding kinetics}\label{app:correlation_time}
{In our analytical model, we assume that the number of binding-unbinding events for each receptor is large. This requires that the time of the observation $T$ is large in comparison to the timescales relevant for binding and unbinding. In this appendix, we determine these timescales. We show that the probability of occupation of a single receptor relaxes to its steady state with two separate timescales $\tau_\textrm{corr}^\pm$, where the correlation time depends on the kinetics of binding (rates $k$, $r$, and $r'$) and the effective concentrations $c$ and $c'$.} With the presence of the incorrect ligands, a receptor can be in an unbound state, a correct bound state, or an incorrect bound state. Thus we can write the corresponding three-state master equations:
\begin{eqnarray}
    \dfrac{\dd p_\text{correct}}{\dd t}&=&(1-p_\text{correct}-p_\text{incorrect})kc-p_\text{correct}r,\nonumber\\
    \dfrac{\dd p_\text{incorrect}}{\dd t}&=&(1-p_\text{correct}-p_\text{incorrect})kc'-p_\text{incorrect}r',\nonumber
\end{eqnarray}
where $1-p_\text{correct}(t)-p_\text{incorrect}(t)$ is probability that the receptor is unbound.
The steady-state solutions can be easily obtained by setting the time derivatives to zero:
\begin{eqnarray}
    &p_\textrm{correct}^{\textrm{SS}}&=\frac{kcr^{-1}}{1+kcr^{-1}+kc'r'^{-1}},\\
    &p_\textrm{incorrect}^{\textrm{SS}}&=\frac{kc'r'^{-1}}{1+kcr^{-1}+kc'r'^{-1}}.
\end{eqnarray}
If there is only one species of ligands, the occupancy (aka fractional saturation) is given by
$p=c/(c+K_D)=kcr^{-1}/(1+kcr^{-1})$,
where $K_D=r/k$ is the dissociation constant~\cite{PhysRevLett.105.048104}. With the presence of the second species of ligands, the {probability of being bound at all} is~\cite{mora}:
\begin{equation}\label{eq:occupancy}
p=p_\textrm{correct}^{\textrm{SS}}+p_\textrm{incorrect}^{\textrm{SS}}=\frac{kcr^{-1}+kc'r'^{-1}}{1+kcr^{-1}+kc'r'^{-1}}.
\end{equation}
However, in order to calculate the receptor correlation timescale, we have to solve the master equations over time. Because the final solutions should converge to the steady state, we assume the complete solutions are in the form of
\begin{eqnarray*}
    p_\textrm{correct}(t)&=&A\ee^{-t/\tau_1}+B\ee^{-t/\tau_2}+p_\textrm{incorrect}^{\textrm{SS}},\\
    p_\textrm{incorrect}(t)&=&C\ee^{-t/\tau_1}+D\ee^{-t/\tau_2}+p_\textrm{incorrect}^{\textrm{SS}},
\end{eqnarray*}
where $p_\textrm{correct}\to p_\textrm{correct}^\textrm{SS}$ and $p_\textrm{incorrect}\to p_\textrm{incorrect}^\textrm{SS}$ when $t\to\infty$.
{Here we are primarily interested in the two correlation times $\tau_1$ and $\tau_2$ -- which shouldn't depend on the particular initial conditions. We then assume the receptor is bound to a correct ligand at $t=0$, so the initial conditions are $p_\textrm{correct}(0)=1$ and $p_\textrm{incorrect}(0)=0$, i.e., $A+B=1-p_\textrm{correct}^{\textrm{SS}}$ and $C+D=-p_\textrm{incorrect}^{\textrm{SS}}$.} Substituting the solutions back to the master equations, we can find the four coefficients:
\begin{eqnarray}
    &A=&-\frac{kc}{2\sqrt{\Delta}}p_\textrm{incorrect}^\textrm{SS}+\frac{1}{2}(1-p_\textrm{correct}^\textrm{SS})\nonumber\\
    &&~~~~~~~~+\frac{(kc+r)-(kc'+r')}{2\sqrt{\Delta}}(1-p_\textrm{correct}^\textrm{SS}),\nonumber\\
    &B=&\frac{kc}{2\sqrt{\Delta}}p_\textrm{incorrect}^\textrm{SS}+\frac{1}{2}(1-p_\textrm{correct}^\textrm{SS})\nonumber\\
    &&~~~~~~~~-\frac{(kc+r)-(kc'+r')}{2\sqrt{\Delta}}(1-p_\textrm{correct}^\textrm{SS}),\nonumber\\
    &C=&\frac{1}{\tau_2}\frac{p_\textrm{incorrect}^\textrm{SS}}{\sqrt{\Delta}},~~~~~~D=-\frac{1}{\tau_1}\frac{p_\textrm{incorrect}^\textrm{SS}}{\sqrt{\Delta}},\nonumber
\end{eqnarray}
and the two characteristic times:
\begin{eqnarray}
    &&\tau_1=\frac{2}{kc+r+kc'+r'+\sqrt{\Delta}},\\
    &&\tau_2=\frac{2}{kc+r+kc'+r'-\sqrt{\Delta}},
\end{eqnarray}
where the discriminant-like term $\Delta = [k(c+c')+r-r']^2-4kc'(r-r')$
is larger than $0$ under our assumption $r'\geqslant r$. We can also rewrite $\Delta$ in a more symmetric form:
\begin{equation*}
    \Delta = {(kc+r+kc'+r')^2-4rr'(1+kcr^{-1}+kc'r'^{-1})},
\end{equation*}
which ensures $\tau_1,\tau_2>0$. Moreover, when there is only one species of ligands, i.e., the correct ligand concentration $c=0$, the two timescales degenerate to $\tau_2=1/r$ and $\tau_1=1/(kc'+r')$, which is exactly the correlation timescale for one species of ligands~\cite{Apara}.
{In brief, we can write the characteristic correlation times $\tau_1$ and $\tau_2$ as
\begin{equation}\label{eq:t_corr}
    \tau_\textrm{corr}^\pm=\frac{2}{kc+r+kc'+r'\pm\sqrt{\Delta}},
\end{equation}
and expect that for our semi-analytical calculation of the Fisher information to be valid, we must be in the limit $T \gg \tau^\pm_\textrm{corr}$ for all receptors.}

\section{Semi-analytical calculation of Fisher information}\label{app:fisher}

If the correct ligand concentration takes the form of $c(x)=c_0 g(x)$, and there are $N$ independent receptors evenly distributed in the cell perimeter $L$, the $n$-th receptor will have a different value of the ephrin concentration $c(x_n)$ and a different number of binding-unbinding events $M_n$ during a measuring time $T$. Then for the $n$-th receptor, the probability to have a time record $\{\tau_b^{(n)},\tau_u^{(n)}\}$ is 
\begin{equation}
P_n=\prod_{i=1}^{M_n}P(\tau_{u,i}^{(n)})\cdot P(\tau_{b,i}^{(n)}),
\end{equation}
where the unbinding and binding times are given by 
\begin{eqnarray}
    &&P(\tau_u)=k(c+c')\ee^{-k(c+c')\tau_u},\label{eq:p_tu_appendix}\\
    &&P(\tau_b)=\frac{c}{c+c'}r\ee^{-r\tau_b} + \frac{c'}{c+c'}r'\ee^{-r'\tau_b},\label{eq:p_tb_appendix}
\end{eqnarray}
Therefore, given the three parameters $\boldsymbol{\theta}=(c_0, c', x_0)$ in our model, the total probability of all $N$ receptors is $P(\{\tau\}|\boldsymbol{\theta})=\prod_{n=1}^N P_n$,
then the likelihood $\mathcal{L}(\boldsymbol{\theta}|\{\tau\})=P(\{\tau\}|\boldsymbol{\theta})$ is given by 
\begin{widetext}
\begin{eqnarray}
    \mathcal{L}&=&\prod_{n=1}^N\prod_{i=1}^{M_n}\left[k(c+c')\ee^{-k(c+c')\tau_{u,i}^{(n)}}\times\left(\frac{c}{c+c'}r\ee^{-r\tau_{b,i}^{(n)}}+\frac{c'}{c+c'}r'\ee^{-r'\tau_{b,i}^{(n)}}\right)\right]\nonumber\\
    &=&\prod_{n=1}^N\ee^{-k(c+c')T_{u}^{(n)}}\prod_{i=1}^{M_n}k\left(rc\ee^{-r\tau_{b,i}^{(n)}}+r'c'\ee^{-r'\tau_{b,i}^{(n)}}\right)\nonumber\\
    &=&\prod_{n=1}^N\ee^{-kc_t[\chi g(x_n) +1-\chi]T_{u}^{(n)}}\prod_{i=1}^{M_n}kr'c_t\ee^{-r'\tau_{b,i}^{(n)}}\left[1-\chi+\alpha\chi g(x_n)\ee^{(1-\alpha)r'\tau_{b,i}^{(n)}}\right]\nonumber\\
    &=&\prod_{n=1}^N\ee^{-kc_t[\chi g(x_n) +1-\chi]T_{u}^{(n)}-r'T_b^{(n)}}\prod_{i=1}^{M_n}kr'c_t\left[1-\chi+\alpha\chi g(x_n)\ee^{(1-\alpha)r'\tau_{b,i}^{(n)}}\right],\nonumber
\end{eqnarray}
\end{widetext}
{where in the second and last steps we have defined the total amount of unbound and bound time for the $n$-th receptor as $T_u^{(n)} = \sum_{i=1}^{M_n} \tau_{u,i}^{(n)}$ and $T_b^{(n)} = \sum_{i=1}^{M_n} \tau_{b,i}^{(n)}$, respectively. And in the penultimate step, we introduce a transformation from $(c_0, c')$ to the total concentration $c_t=c_0+c'$ and correct ligand fraction $\chi=c_0/c_t$, i.e., we substitute $c'=c_t(1-\chi)$, $r=\alpha r'$, and $c(x)=c_t\chi g(x)$.} 
After taking a logarithm, $\ln\mathcal{L}$ can be written as a sum of five independent contributions:
\begin{eqnarray}
    &&\ln\mathcal{L}_0=\sum_{n=1}^N\sum_{i=1}^{M_n}\ln kr'=\sum_{i=1}^{N}M_n\ln k r',\nonumber\\
    &&\ln\mathcal{L}_1=-r'\sum_{n=1}^N\sum_{i=1}^{M_n}\tau_{b,i}^{(n)}=-r'\sum_{n=1}^NT_b^{(n)},\nonumber\\
    &&\ln\mathcal{L}_2=\sum_{n=1}^N\sum_{i=1}^{M_n}\ln c_t=\sum_{n=1}^N M_n \ln c_t,\label{eq:log-likelihood}\\
    &&\ln\mathcal{L}_3=\sum_{n=1}^N\sum_{i=1}^{M_n}\ln\left[1-\chi+\alpha\chi g(na)\ee^{(1-\alpha)r'\tau_{b,i}^{(n)}}\right],\nonumber\\
    &&\ln\mathcal{L}_4=-\sum_{n=1}^N k c_t\left[1-\chi+\chi g(na)\right]T_u^{(n)},\nonumber
\end{eqnarray}
where $a$ is the receptor spacing and $x_n=na$ gives the position of the $n$-th receptor. Note that $\mathcal{L}_0$ and $\mathcal{L}_1$ are trivial terms since they don't depend on any parameter we are concerned with, and thus we only need to include the other three terms in the following computation. 
Calculating the derivatives $\partial^2\ln\mathcal{L}/\partial\theta_\alpha\partial\theta_\beta$ is straightforward but not particularly informative. We include these results in a Mathematica notebook.
These derivatives depend on the bound/unbound times through terms like $T_u^{(n)}, M_n$, but also in more complex forms $f(\tau_{b,i}^{(n)})$ like
$\ee^{(1-\alpha)r'\tau_{b,i}^{n}}$.

The complex part during the derivation of the Fisher information matrix $I(\boldsymbol{\theta})_{\mu\nu}=-\langle\partial^2\ln\mathcal{L}/\partial\theta_\mu\partial\theta_\nu\rangle$ arises from computing the average $\langle\cdots\rangle$. To begin, we can calculate the average of the bound and unbound times for the $n$-th receptor:
\begin{eqnarray}\label{eq:avg}
&&\langle\tau_{u}^{(n)}\rangle=\int_0^\infty\tau P(\tau_u^{(n)} =\tau)\dd\tau=\frac{1}{k(c_n+c')},\nonumber\\
&&\langle\tau_{b}^{(n)}\rangle=\int_0^\infty\tau P(\tau_b^{(n)} =\tau)\dd\tau=\frac{c_n}{r(c_n+c')}+\frac{c'}{r'(c_n+c')}.\nonumber
\end{eqnarray}
We then assume the average of the total binding-unbinding events $M_n$ during the measuring time $T$ can be approximated as
\begin{equation}
    \langle M_n\rangle = \frac{T}{\langle\tau_{u}^{(n)}\rangle + \langle\tau_{b}^{(n)}\rangle},
\end{equation}
which should only be valid when $T\gg\tau_\textrm{corr}^\pm$ (Appendix \ref{app:correlation_time}). And the average of the total unbound time for the $n$-th receptor is
\begin{equation}
    \langle T_u^{(n)}\rangle=(1-p_n)T,
\end{equation}
where $p_n$ is the occupancy given by Eq. \eqref{eq:occupancy}.
We can readily verify $\langle M_n\rangle = {\langle T_u^{(n)}\rangle}/{\langle \tau_u^{(n)}\rangle}$, which is expected to be true if $T\gg\tau_\textrm{corr}^\pm$. When computing a sum over all binding-unbinding events, {e.g., $ \langle\sum_{i=1}^{M_n} f(\tau_{b,i}^{(n)}) \rangle$, we assume that the average of each term is equal, and as a result, it's equivalent to multiply the average of a single term by $\langle M_n\rangle$. This essentially treats the value of the number of binding-unbinding events $M_n$ as fixed, and requires again that we have $T \gg \tau_\textrm{corr}^\pm$.} {To sum up, to go from the analytical derivatives of Eq. \eqref{eq:log-likelihood} to computing the averages required for the Fisher information matrix, we apply the rules:}
\begin{eqnarray}
    \langle\sum_{n=1}^N M_n\rangle&\to&\sum_{n=1}^N\langle M_n\rangle,\nonumber\\
    \langle\sum_{n=1}^N T_u^{(n)}\rangle&\to&\sum_{n=1}^N \langle T_u^{(n)}\rangle,\\
    \langle\sum_{n=1}^N\sum_{i=1}^{M_n}f(\tau_{b,i}^{(n)})\rangle&\to&\sum_{n=1}^N\langle M_n\rangle\int_0^\infty\dd\tau ~P(\tau_b^{n}=\tau) f(\tau),\nonumber\label{eq:avg_approx}
\end{eqnarray}
where the last term is computed by numerical integration using Gaussian quadrature ({scipy.integrate.quad method provided by SciPy \cite{2020SciPy-NMeth}}). 

{We can see from Eq. \eqref{eq:log-likelihood} that all terms in the Fisher information matrix should contain $\langle M_n\rangle$ or $\langle T_u\rangle$ -- so we expect that the Fisher information time is thus proportional to the total measuring time $T$, at least in the limit of $T\gg\tau_\textrm{corr}$ where our analytic theory is appropriate.}
After obtaining all the terms in the matrix numerically, we can compute its inverse and take the diagonal elements as the lower bound of the variances, namely the Cram\'er-Rao bound in Eq. \eqref{eq:CRB}.

\begin{table*}[htbp]
\caption{\label{tab:param}%
Table of simulation parameters\footnote{These parameters are used throughout the paper; any deviation from them is explicitly noted.}}
\begin{ruledtabular}
\begin{tabular}{llll}
Parameter & Description & Dimension & value\\\hline
$c_0$ & Ephrin concentration & $L^{-2}$ & $10^2\,\mathrm{\mu m}^{-2}$~\cite{xu_epha2_2011}\\
$\chi$ & Fraction of correct ligands $c_0/c_t$ & $1$ & Varying\\
$k$ & Eph-ephrin binding constant & $L^2/T$ & $10^{-2}\,\mathrm{\mu m^2/s}$\\
$r$ & Eph-ephrin unbinding rate & $T^{-1}$ & $1\,\mathrm{s}^{-1}$\\
$\alpha$ & Ratio of unbinding rates $r/r'$ & $1$ & Varying\\
$x_0$ & Contact center position & $L$ & $50\,\mathrm{\mu m}$ \\
$\sigma$ & Contact width & $L$ &  Varying\\
$\omega$ & Boundary steepness & $L$ & $1\,\mathrm{\mu m}$\\ 
$T$ & Measuring time & $T$ & $1\,\mathrm{s}$\\
$L$ & Cell perimeter & $L$ & $100\,\mathrm{\mu m}$ \\
$a$ & Eph receptor spacing & $L$ & $0.01\,\mathrm{\mu m}$ \\
$N$ & Number of receptors & $1$ & $10000$\\
\end{tabular}
\end{ruledtabular}
\end{table*}

\section{Details of stochastic simulations}\label{app:stochastic}

\subsection{Simulation methods}

In the Monte Carlo simulations, we generate a series of random bound/unbound times {for each receptor} from the probability distribution in Eq. \eqref{eq:p_tu} and \eqref{eq:p_tb}.
To do this, we note that Eq. \eqref{eq:p_tb} is essentially a composite of two exponential distributions: we draw samples from $r\ee^{-r\tau_b}$ with a probability of $c/(c+c')$ and from $r'\ee^{-r'\tau_b}$ with a probability of $c'/(c+c')$, which is equivalent to running a kinetic Monte Carlo simulation to generate these time series.
Subsequently, we compute the log-likelihood function $\ln\mathcal{L}(\boldsymbol{\theta}'|{\tau})$ using Eq. \eqref{eq:log-likelihood} and numerically maximize it to find the optimal parameters $(\hat{c}_t,\hat{\chi},\hat{x}_0)$.
When doing the optimization, we use the modified Powell algorithm~\cite{powell1964efficient} (provided by SciPy~\cite{2020SciPy-NMeth}) to minimize $-\ln\mathcal{L}$ within the bounds $c_t\in[0, 10^{8}]\,\mathrm{\mu m}^{-2},~\chi\in[0, 1],~x_0\in[0, 100]\,\mathrm{\mu m}$, and set the relative error in solutions acceptable for convergence to $10^{-7}$. {This gives us the maximum-likelihood estimators $(\hat{c}_t,\hat{\chi},\hat{x}_0)$ for this individual trajectory. We then repeat this process many times for new trajectories ($\sim 10^4$), thus allowing us to compute distributions of the estimators. We then plot the standard deviations of these estimator distributions in, e.g., Fig. \ref{fig:var}}.

We have not been able to compute the false negative rates in Eq. \eqref{eq:FNR} and false positive rates in Eq. \eqref{eq:FPR} analytically. However, we can still calculate the FNR and FPR through Monte Carlo simulations. To compute FNR, we assume the other cell is present, i.e., $c_0\neq 0$, and then generate a time series to compute the likelihood ratio {by computing the maximum log-likelihood (Eq. \ref{eq:log-likelihood}) numerically under both the assumption of the other cell being present and it being absent. We then repeat this process many times to compute the FNR for a fixed value of $\lambda$.} For the FPR, we do simulations by first assuming the other cell is absent (setting $c_0=0$), but keep all other parameters the same. We perform approximately $\sim 10^5$ simulations to do averages for each data point shown in Fig. \ref{fig:FNR}, which allows us to obtain robust and reliable results for the analysis.

\subsection{Parameters}
The default parameter values in our model are shown in Table \ref{tab:param}. The typical radius of the \emph{Drosophila} haemocytes cell is in the order of $10\,\mathrm{\mu m}$~\cite{exp}, which makes the cell perimeter roughly $10^2\,\mathrm{\mu m}$. The concentration of EphA2 receptors on the cell surface is reported to be around $100-1000/\mathrm{\mu m}^2$ for CHO cells~\cite{https://doi.org/10.1002/syst.202200011} and about $600/\mathrm{\mu m}^2$ for A549 lung cancer cells~\cite{singh_epha2_2018}. As we're modeling the perimeter of the cell as a line, we set the contact center $x_0=L/2$ to avoid boundary effects. Additionally, we choose a receptor spacing of $a=0.01\,\mathrm{\mu m}$ to ensure that there are approximately $100$ receptors per square micron of the surface area on the cell, assuming a height in the $z$-direction around the order of $1\,\mathrm{\mu m}$.

\subsection{Consistency between stochastic simulation and semi-analytic theory}

The fundamental assumption we utilize when deriving the Fisher information matrix in Appendix \ref{app:fisher} is that $T\gg\tau_\textrm{corr}^\pm$ for all receptors on the cell. {In this limit, it is reasonable to compare our theory, which has a fixed number of binding/unbinding events for each receptor, to the stochastic simulations, where we fix the measuring time $T$. 
This assumption leads to the theoretical Cram\'er-Rao bound scaling as $\sim 1/T$ (Appendix \ref{app:fisher}). When we compare simulation and theory, we should check this dependence, to ensure we are exploring times $T\gg\tau_\textrm{corr}^\pm$. We find that the variances obtained from simulations don't depend strongly on $T$ when $T\ll$ 1 second and the discrepancy between simulation and theory grows with decreasing measuring time (Fig. \ref{fig:T_scale}). This makes sense: in the limit $T \ll \tau_\textrm{corr}$, the cell is effectively making a ``snapshot'' measurement, using the receptor state at one instant to estimate parameters, and the error in measurement will not depend strongly on the measurement time. The correlation timescale of about a second is consistent with our analysis in Appendix \ref{app:correlation_time}. 
With the chosen binding rate $kc$ and unbinding rate $r$ in the order of $1\,\mathrm{s^{-1}}$ (Table \ref{tab:param}), the correlation times $\tau_\textrm{corr}$ given by Eq. \eqref{eq:t_corr} are $\tau_1\sim10^{-1}\,\mathrm{s}$ and $\tau_2\sim1\,\mathrm{s}$. Therefore, to effectively compare the simulation results with the theoretical Cram\'er-Rao bound in Eq. \eqref{eq:CRB}, we need to increase the measuring time to $T\sim10\,\mathrm{s}$ in Fig. \ref{fig:var} and Fig. \ref{fig:Gaussian}.}

\newpage

\setcounter{figure}{0}
\renewcommand{\thefigure}{S\arabic{figure}}

\begin{figure*}[htbp]
\includegraphics[width=0.8\textwidth]{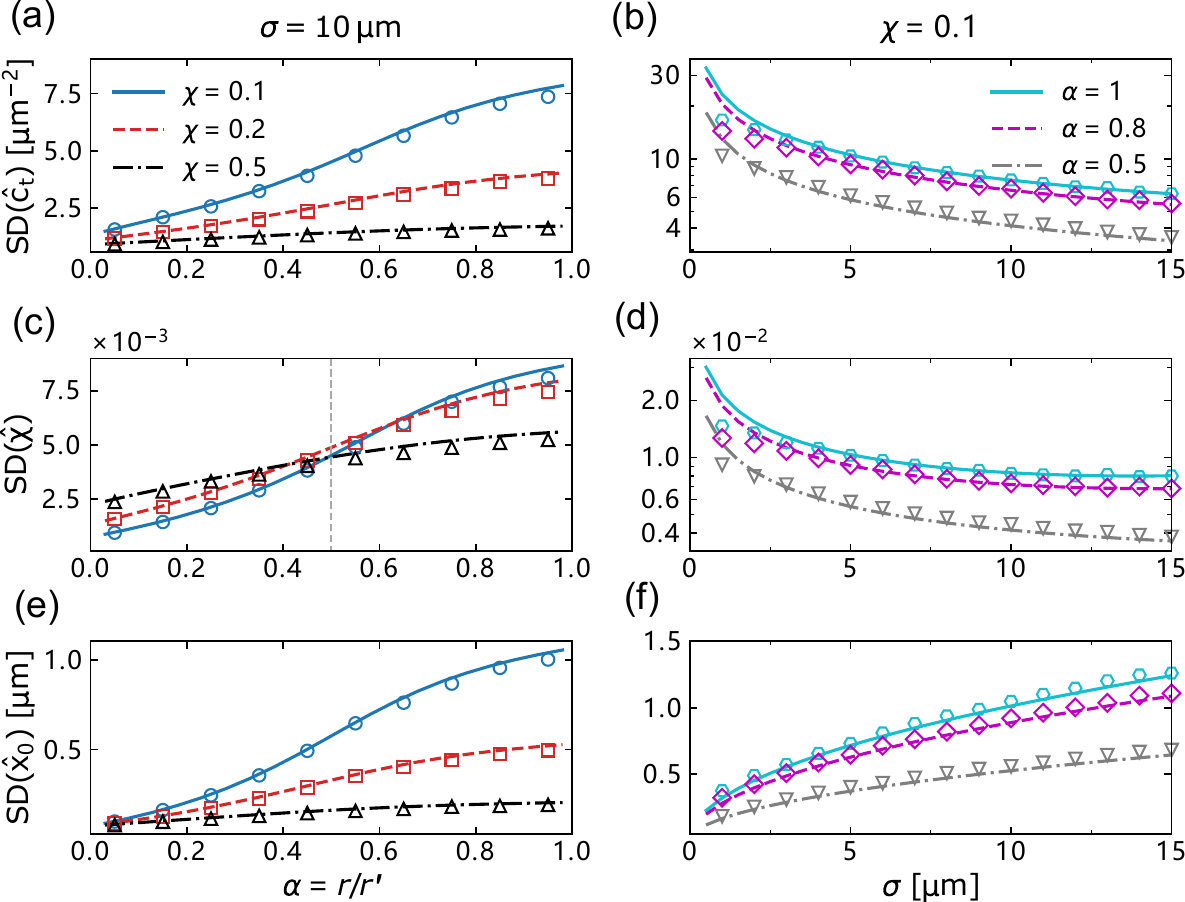}%
\caption{\label{fig:Gaussian} 
Estimating errors of MLE when the correct ligand concentration $c(x)$ takes the form of a Gaussian function  $c(x)=c_0\exp(-(x-x_0)^2/2\sigma^2)$ rather than the rectangular function given in Eq. \eqref{eq:cx}. (a-b) show the standard deviation (SD) of the total concentration $\hat{c}_t$, (c-d) show the SD of correct ligand fraction $\hat{\chi}$, and (e-f) display the SD of contact center position $\hat{x}_0$, where the estimating error of $x_0$ increases as the contact width $\sigma$ expands. In all figures, the lines (solid, dashed, and dash-dot) represent theoretical results obtained from  Eq. \eqref{eq:CRB}, while the points are obtained from Monte Carlo simulations (the standard deviations are computed from $10^4$ simulations for each point); Measuring time is $T=10\,\rm{s}$.}
\end{figure*}

\begin{figure*}[htbp]
\includegraphics[width=0.4\textwidth]{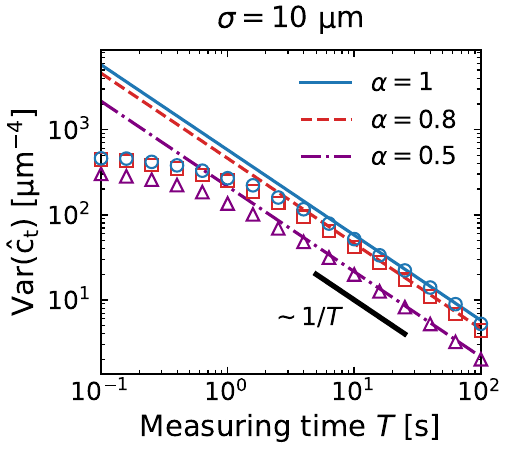}%
\caption{\label{fig:T_scale}
Variances of MLE decrease with increasing measuring time $T$.
The variances from the Cram\'er-Rao bound scales as $\sim 1/T$ in our derivation (Appendix \ref{app:fisher}). Though the simulation results show a slower decrease compared to the theoretical predictions at small $T$, they still converge as the measuring time $T\gg\tau_\textrm{corr}$. Lines represent theoretical results obtained from  Eq. \eqref{eq:CRB}, while the points are obtained from Monte Carlo simulations; Correct ligand fraction is $\chi=0.1$.}
\end{figure*}

\begin{figure*}[htbp]
\includegraphics[width=0.4\textwidth]{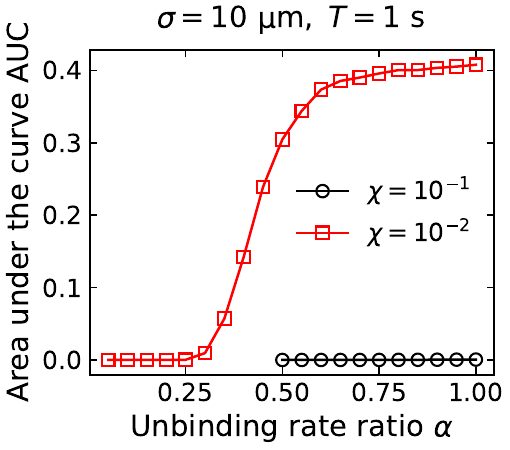}%
\caption{\label{fig:chi_AUC}
Area under the DET curve changing with unbinding rate ratio $\alpha$ when the correct ligand fraction $\chi=10^{-2}$. Black points are exactly the results in Fig. \ref{fig:ROC} (e) when $\chi=10^{-1}$.}
\end{figure*}

\end{document}